# The growth mechanism of CuO nanowires synthesized by the thermal oxidation method

Lilin Xie[a], Yoshifumi Oshima[a], and Xiaona Zhang[b*]

a. School of Materials Science, Japan Advanced Institute of Science and Technology, Ishikawa 923-1292, Japan.

b. Institute of Microstructure and Property of Advanced Materials, Beijing University of Technology, Beijing 100124, People's Republic of China

**Abstract**

Large-scale CuO nanowires are commonly synthesized by the thermal oxidation method. However, the growth mechanism remains controversial between diffusion growth mechanism and vapor-solid (VS) growth mechanism. In this study, we investigated the morphology and atomic structure of the CuO nanowires by electron microscopes to clarify the growth mechanism. We found that they were grown with the annealing ambient during the cooling process and there were screw dislocations in the nanowires. These features indicate that the VS growth mechanism is applicable to explain the growth of the CuO nanowires. In addition, the kinetic consideration also supports the VS growth mechanism.

## 1. Introduction

The one-dimensional (1D) materials, such as nanowires or nanorods, have attracted great research interests. Owing to their low dimensional structures, nanowires present various novel physical properties, such as excellent mechanical properties with high strength and plasticity[1], electrical properties with ballistic transport and quantized conductivity[2]. And they have been demonstrated to be applicable to nano-devices, such as the nanogenerator[3] and optoelectronic devices[4,5]. As a representative metal oxide, copper oxide (CuO) is a typical p-type semiconductor with a bandgap of 1.2 eV[6]. It has been widely investigated due to their promising applications in catalysis[7], field-effect transistor[8], nano-photodetector[9], and copper lithium oxide battery[10]. Therefore, a convenient, low-cost, and high-crystallinity method for the large-scale preparation method has been demanded.

The thermal oxidation method[11,12] is a simple and commonly used method to synthesize CuO nanowires, in which the large-scale nanowires with high crystallinity can be synthesized on the Cu surface by oxidization. The growing conditions and mechanism of CuO nanowires have been extensively studied[11,13,14]. CuO nanowires have been reported to grow at a temperature range of 400-750 ℃ in the oxygen atmosphere or the air ambient. However, the growth mechanism of CuO nanowires has been still ambiguous such that two possible growth mechanisms, vapor-solid (VS) growth mechanism[15–18] and diffusion growth mechanism[14,19,20], have been proposed. In the VS growth mechanism, the gaseous atoms are adsorbed at the front, causing the nanowires to grow layer-by-layer in one direction[21–26], which is similar to whisker growth[27,28]. Such a 1D growth is contributed by the screw dislocation, which could provide perpetually atomic steps that promote the deposition of vapor[22]. Therefore, the screw dislocation plays a crucial role in the VS growth mechanism. Additionally, the VS growth mechanism has the characteristic that the size of nanowires seems to control easily because the width is determined in the initial state and does not change in the subsequent process. Since it uses gaseous atoms and molecules at high growth temperature, the growth rate of nanowires is relatively fast. The growth mechanism of metal oxides nanowires[11,29–31] such as $Fe_2O_3$ and tungsten oxide have been explained by VS growth mechanism.

On the other hand, in the diffusion growth mechanism, the nanowires are grown by stress-induced diffusion[13,19,32] or short-circuit/grain boundary diffusion[14,33,34] of Cu atoms from the substrate to the CuO grain. They have been suggested to be driven by interfacial stress origin from the mismatch between the lattice structure of CuO (monoclinic with $a$ = 4.684 Å, $b$ = 3.425 Å, $c$ = 5.129 Å, and $\beta$ = 99.47°) and $Cu_2O$ (cubic with $a$ = 4.22 Å.)[13], and the diffusion rate of Cu atoms/ions in the twin/grain boundary is higher than that of the inside, resulting in the one-dimensional growth. While, such a diffusion growth could not explain several experimental results[13] that the density of nanowires was reduced with gas flow, and both density and diameter were not influenced by annealing time. It is conceivable that the difference in the growth mechanism will result in different structures. Therefore, to clarify the growth mechanism of CuO nanowires, further investigation on the crystal structure should be carried out.

In this work, we have observed the CuO nanowires, which were synthesized by the thermal oxidation method, through the scanning electron microscope (SEM) and transmission electron microscope (TEM) observation to clarify the growth mechanism. The CuO nanowires are found to grow during the cooling process and have a specific structural feature of screw dislocation. With further consideration of the growth kinetic, we suggest that the growth of CuO nanowire shall be attributed to the VS growth mechanism.

## 2. Experimental method

The CuO nanowires were synthesized by the thermal oxidation of the Cu substrate, which was a standard TEM Cu grid. The Cu substrate was firstly cleaned with an aqueous of 1.0 mol/L HCl solution for 10 s, and then rinsed with distilled water. After drying with air flow, the cleaned substrate was placed on an alumina board and put into a muffle furnace in the atmosphere of air. During the thermal oxidation process, the temperature was firstly increased to the annealing temperature (550°C) with a heating rate of 10 °C/min, then it was maintained at to the annealing temperature for the annealing time of 5 hours. After the annealing, the samples were cooled to room temperature with keeping the annealing ambient or changing the ambient to fresh air.

To investigate the morphology of CuO nanowires, the samples were peeled from the substrate and observed by SEM (FEI Quanta 250, 20 kV). For the structural analysis, the CuO nanowires were scraped from the substrate, sonicated in acetone, and dropped to the TEM grid with holey carbon films to achieve the TEM (JEOL-2010, 200 kV) and high-resolution TEM (HR-TEM, JEOL-2010F, 200 kV) observation.

## 3. Experimental results

### *3.1 The morphology of CuO nanowires*

Figure 1(a) shows the cross-section SEM image of the CuO nanowires grown on the Cu substrate with the annealing temperature of 550 °C. The CuO nanowires were found to grow on the $Cu_2O$ and CuO layers, which were sequentially formed on the substrate. Such a stacking was consistent with the previous reports[12,14,19] which showed that the Cu substrate was oxidized in two steps[35–37]: it was firstly oxidized to form $Cu_2O$, and then $Cu_2O$ was further oxidized to form CuO. The CuO nanowires were also grown with the formation of the CuO layer. In the SEM image, the CuO nanowires were observed to be formed at high density and their diameters were estimated to be about 100 nm on average and the length was 5 to 15 μm.

The $Cu_2O$ and CuO layers were grown due to thermal oxidation of the Cu substrate, however, the growth mechanism of the CuO nanowires has been controversial as mentioned above. In the VS mechanism, the atomic step acts as a growth site by facilitating the adsorption of gaseous Cu and O atoms, similar to whisker growth, as shown in Fig. 1 (b). On the other hand, in the diffusion growth mechanism, the Cu atoms/ions are diffused along the grain/twin boundaries to form CuO nanowires, as shown in Fig. 1 (c). Such nanowires seem to have a shape surrounded by stable surfaces and grow in a fixed direction. Therefore, the structure of the CuO nanowires is thought to depend on the growth mechanism. But, there are no reports that clarify the structure of the CuO nanowires.

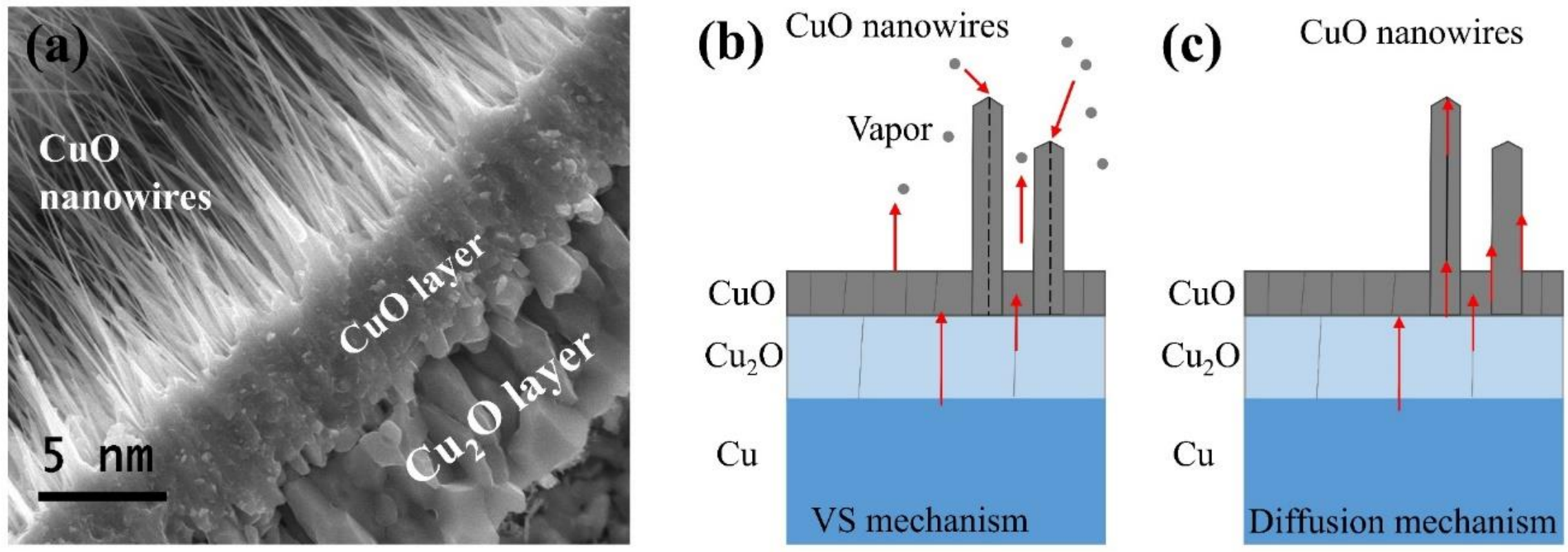

Figure.1. (a) SEM image of CuO nanowires synthesized by thermal oxidation method. (b)The diagram of the VS growth mechanism, which is contributed to the deposition of vapor promoted by the screw dislocation. (c)The diagram of the diffusion growth mechanism, which is contributed to the diffusion of atoms/ions along the twin/grain boundaries. The red arrows indicate the diffusion and deposition paths.

Meanwhile, we noticed that the annealing ambient was crucial on the growth of CuO nanowires, as shown in Fig. 2. It was found that the CuO nanowires could only grow by keeping the substrate with the annealing ambient during cooling, as shown in Fig. 2(a). While, the CuO nanowires would not grow by changing the annealing ambient to fresh air during the cooling process, only the outcrop of nanowires/nanorods have been found, as shown in Fig. 2(b). It means that the CuO nanowires were grown during the cooling process with the annealing ambient, such a result has not been clarified in previous reports. Obviously, the diffusion growth mechanism is not suitable for explaining the observed results, since the diffusion of Cu atoms/ions would be suppressed during the cooling process. Therefore, the growth of CuO nanowires should be attributed to the vapor-deposition growth, because the annealing ambient is essential, and the cooling can contribute to the deposition of vapor. To confirm it, further structural analysis on the CuO nanowires is needed.

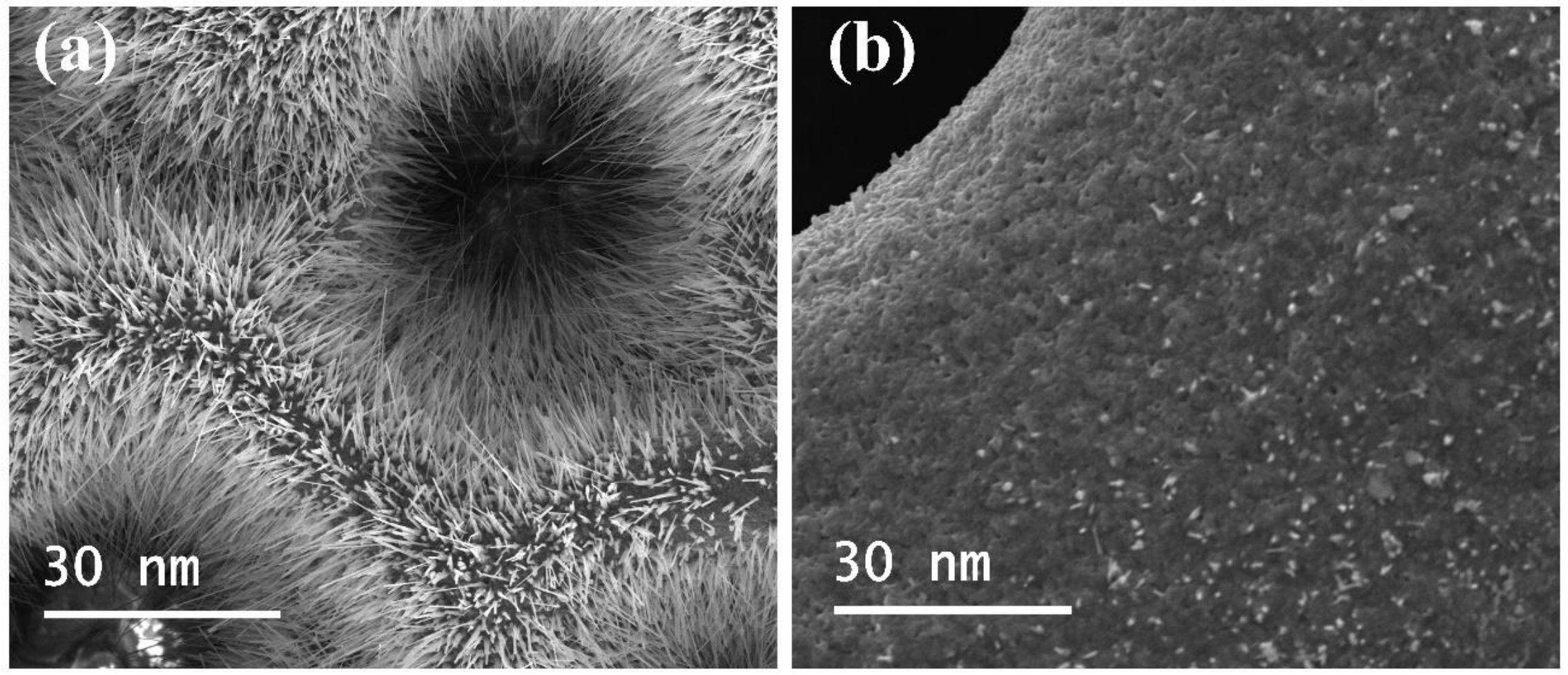

Figure.2. SEM images of CuO nanowires synthesized at different cooling conditions. (a) Cooling with annealing ambient of the furnace. (b) Cooling with fresh air.

### *3.2 The screw dislocation in the CuO nanowires*

Figures 3(a) and (b) show typical TEM images of the CuO nanowires. The selected area electron diffraction (SAED) inserted in Fig.3 (a) shows the [001] zone axis pattern (ZAP) of the CuO crystal. According to the TEM image and the SAED pattern, the growing direction of the nanowire was determined to be the [110] direction. A black line along the axis of the nanowire, as shown in both Fig.3 (a) and (b), was thought to correspond to a dislocation line. In addition, several dark-belt, which were formed as indicated by an arrow in Fig.3 (b), was thought to correspond to Eshelby twist[38] as previously reported[39,40]. Since Eshelby twist has been reported to be formed when the screw dislocation existed in the nanowire[28,41], the dark line observed in the CuO nanowires may be a screw dislocation line. Considering that the Eshelby twist has a period depending on the growth condition[42], the period of the corresponding pattern would be different among the nanowires, and sometime such pattern did not appear clearly as shown in Fig. 3(a).

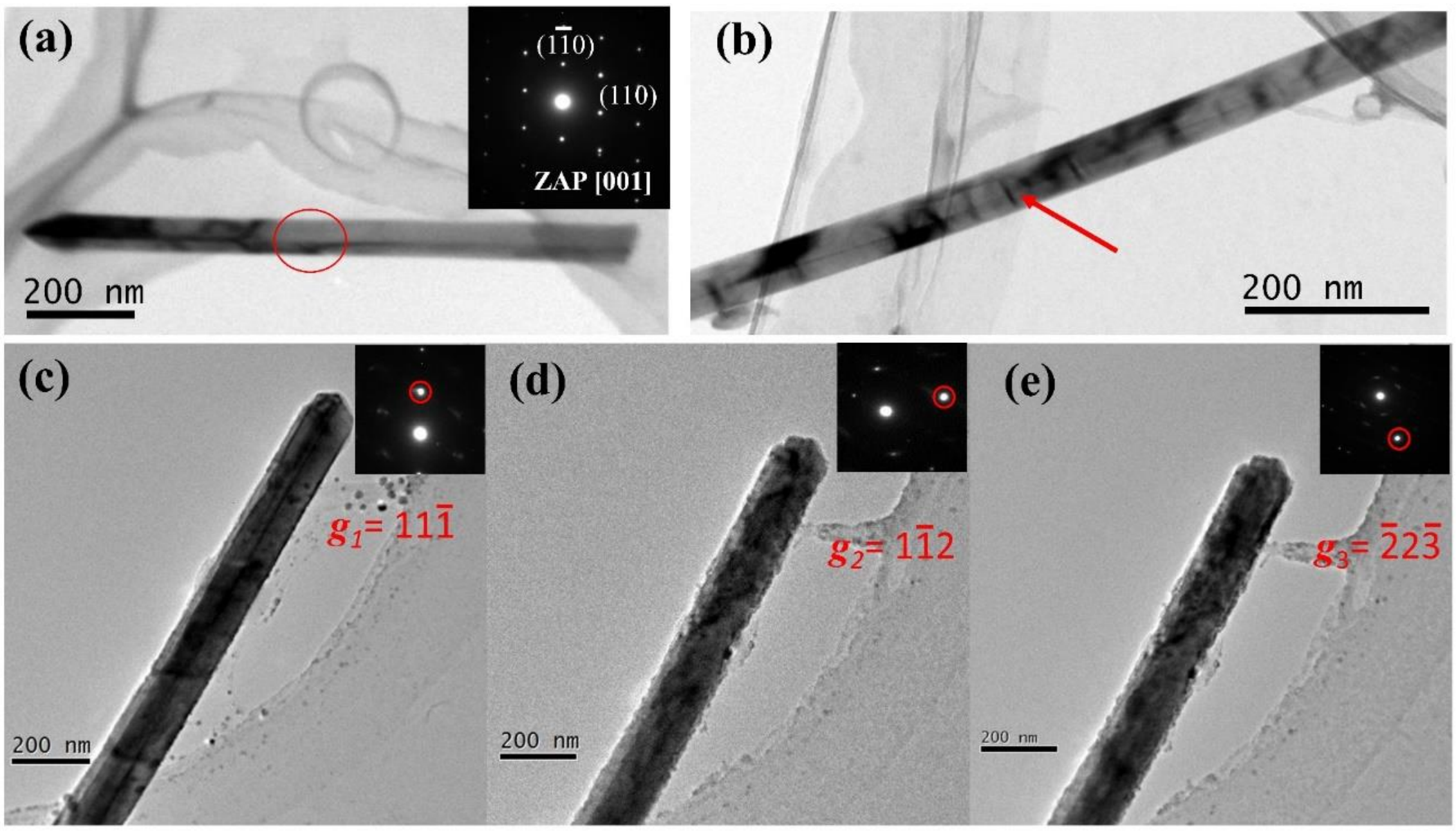

Figure.3. The typical TEM image of CuO nanowires. (a) The TEM image shows a dislocation contrast, inset is the diffraction pattern with a [001] ZAP of CuO. (b)The TEM image shows the contrast of dislocation and the Eshelby twist is indicated by the arrow. (c)-(e) TEM images of a CuO nanowire under two-beam condition, insets are the corresponding diffraction patterns with transmission beam and diffraction beam (marked with a red circle and showed with corresponding *g* vectors). Dislocation contrast is visible in (c), while invisible in (d) and (e).

To confirm the type of dislocation, we performed the TEM observation under two-beam condition and determined the Burges vector. The two-beam condition was achieved by tilting the crystal to excite a particular diffraction spot, ***g*** (hkl), more strongly than the other diffraction spots. When the Burgers vector, ***b***, is perpendicular to ***g***, such as $\boldsymbol{g} \bullet \boldsymbol{b} = 0$, the contrast corresponding to the dislocation disappears. The TEM images of Figs. 3 (c)-(e) were taken by exciting different diffraction spots. The dislocation contrast was visible in Fig. 3(c) when exciting the diffraction spot $\boldsymbol{g_1}$ (11-1), while invisible in Figs. 3(d) and (e) when exciting the diffraction spots, $\boldsymbol{g_2}$ (1-1-2) and $\boldsymbol{g_3}$ (-223), respectively. Due to the invisibility criterion, the Burgers vector of the dislocation, obtained by taking the cross product of those two invisible vectors $\boldsymbol{g_2}$ and $\boldsymbol{g_3}$: $\boldsymbol{b} = \boldsymbol{g}_2 \times \boldsymbol{g}_3$, was determined to be [110]. The [110] Burgers vector was also confirmed in another nanowire (see supplemental materials for the details). These results indicate that a screw dislocation is formed in the CuO nanowire, since the Burgers vector $\boldsymbol{b}$ is parallel to the dislocation line (parallel to the growth direction). It suggests that the 1D growth of the CuO nanowires is driven by the screw dislocation and should be explained by the VS growth mechanisms.

## 4. Discussion: the growth mechanism of CuO nanowire on the perspectives of growth kinetics

In this work, the growth of the CuO nanowires could be explained by the VS growth mechanism instead of the diffusion growth mechanism according to the SEM and TEM analysis on the morphology and atomic structure of nanowires. Because we found vapor is crucial for the growth of CuO nanowire and the screw dislocation in the nanowire could promote the 1D growth by the adsorption of Cu or O atoms from the vapor.

Such VS growth mechanism was investigated by considering the growth kinetics. We found that the density and diameter of nanowires increased with the annealing temperature but were independent of the annealing time (see supplemental materials S2 for the details). Our experimental results agreed with the previous report[13], which showed that the density of the CuO nanowires was independent of the annealing time, the diameters were almost the constant (about 47 nm) in the range of the annealing time from 10 min to 120 min and the long nanowires ( 1.5 μm) were grown by only 5 min at 400 °C. Such growth behaviors could not be explained by the diffusion of Cu atoms/ions, because the diameter should increase due to the atomic diffusion along the nanowire. Moreover, the length should be much short in the case of diffusion growth mechanism due to the short annealing time (5 min) and reduced temperature after annealing. The growth rate of the diffusion growth mechanism [32] can be described as:

$$v = \frac{dL}{dt} = \frac{4DAmP}{(\pi \rho d^2)RTk' t^{1/2}}, \tag{1}$$

where $L$ is the length of the nanowire, $t$ is the time, $D$ is the diffusion coefficient, $A$ is the area, $m$ is the molecular mass of CuO, $P$ is the pressure of oxygen, $d$ is the diameter, $\rho$ is the density of, $R$ is the gas constant, $T$ is the absolute temperature, $k'$ is the rate constant for scaling oxide growth. According to Eq. (1), the relationship between the length and the time follows the power scaling law: $L \sim t^{1/2}$. It is inconsistent with our experiment results that the longest length of the nanowire with the annealing time of 4 hours is about 10 μm, which is close to that of annealing time of 20 hours (see supplemental materials S2 for the details). Furthermore, the driving force in the diffusion mechanism has been suggested to be the interfacial stress between the CuO and $Cu_2O$ crystalline. But, it could hardly affect the large region of several micrometers.

Instead, the VS growth mechanism is more appropriate to explain the observed results. The axial growth rate of the CuO nanowire with the screw dislocation in the VS growth[23] can be calculated as:

$$v = \frac{2\lambda_x p}{\rho d}\left(\frac{m}{2\pi RT}\right)^{\frac{1}{2}}, \tag{2}$$

where $\lambda_x$ is the mean displacement of adsorbed molecules, $p$ is the vapor pressure. Moreover, the radial growth, which determines the change in diameter, increases only when the vapor pressure is larger than the critical vapor pressure $p_0$. Therefore, both the axial and radial growth are controlled by the pressure of vapor. Our experimental results could be explained with such a mechanism. For example, the diameter of CuO nanowires is found to be constant with increasing annealing time. It is because the $p$ is lower than the $p_0$ at a certain annealing temperature, so that there is no radial growth with increasing growth time. Moreover, the diameter of nanowires is found to increase with rising annealing temperature, it is due to the increase in vapor pressure with increasing temperature. Further, the thermal annealing could act as the incubation of CuO vapor, so it is in agreement with the formation of long nanowires in the annealing ambient for a short annealing time. Besides, the reduced temperature during the cooling enhances the deposition rate that increases the growth rate of nanowires, as demonstrated in Eq. (2). Our consideration is also consistent with the consideration in terms of growth rates, the growth rate of the CuO crystal with screw dislocation is $2\lambda_x/d$ times [23] higher than that of without screw dislocation, which is a much higher rate since the $\lambda_x$ is about several micrometers and $d$ is about 100 nm. Therefore, CuO crystals with screw dislocation (VS growth mechanism) grow at a much higher rate and form nanowires. Thus, with the consideration of the growth kinetics, we concluded that the CuO nanowires were formed by the VS growth mechanism.

## 5. Conclusion

In summary, the growth mechanism of the CuO nanowires, which were synthesized by the thermal oxidation method, has been investigated by SEM and TEM observations. We found that the annealing ambient was crucial for the growth of CuO nanowires, and the nanowires grew during the cooling process. With the TEM analysis under two-beam condition, the screw dislocation in the nanowire was confirmed, which suggests a VS growth mechanism for CuO nanowires. Moreover, according to the analysis on the growth kinetics of the CuO nanowires, such as the diameter and the growth rate of nanowires, we find that the VS growth mechanism is more suitable for explaining the experimental results rather than the diffusion growth mechanism. Our results deepen the understanding of the growth of metal oxide nanowires prepared by the thermal oxidation method. It can be expected that the revealed VS growth mechanism could be helpful for growing large-scale and high-crystallinity metal oxide nanowires by a simple thermal oxidation method.

## Declaration of Competing Interest

The authors declare no competing financial interest.

## Acknowledgement